\documentclass[%
 reprint,
superscriptaddress,
 amsmath,amssymb,
 aps,
 pre,
]{revtex4-2}

\usepackage{dcolumn}
\usepackage{bm}
\usepackage{todonotes}
\usepackage{xurl}
\usepackage[hidelinks]{hyperref}
\usepackage[dvipsnames]{xcolor}

\begin{document}


\preprint{APS/123-QED}     


\title{Towards probing velocity distributions in dense granular matter: Utilizing Fiber Bragg Gratings}

\author{Marlo Kunzner}
\affiliation{Institut für Materialphysik im Weltraum, Deutsches Zentrum für Luft- und Raumfahrt (DLR), 51170 Cologne, Germany}
 \altaffiliation[Also at ]{Department for Theoretical Physics, University of Cologne.}
\author{Luis Henriques}
\affiliation{Institut für Materialphysik im Weltraum, Deutsches Zentrum für Luft- und Raumfahrt (DLR), 51170 Cologne, Germany}
 \email{Marlo.Kunzner@dlr.de}
\author{Fahad Puthalath}
\affiliation{Institut für Materialphysik im Weltraum, Deutsches Zentrum für Luft- und Raumfahrt (DLR), 51170 Cologne, Germany}
\author{Leonardo Facchini}
\affiliation{Institut für Materialphysik im Weltraum, Deutsches Zentrum für Luft- und Raumfahrt (DLR), 51170 Cologne, Germany}
\author{Mohammadhossein Shahsavari}
\affiliation{Institut für Materialphysik im Weltraum, Deutsches Zentrum für Luft- und Raumfahrt (DLR), 51170 Cologne, Germany}
\author{Léa Gommeringer}
\affiliation{Institut für Materialphysik im Weltraum, Deutsches Zentrum für Luft- und Raumfahrt (DLR), 51170 Cologne, Germany}
 \altaffiliation[Also at ]{Department for Engineering, University of Munich.}
\author{Martin Angelmahr}
 \affiliation{Fraunhofer Heinrich Hertz Institute (HHI), 38640 Goslar, Germany}
\author{Peidong Yu}
\affiliation{Institut für Materialphysik im Weltraum, Deutsches Zentrum für Luft- und Raumfahrt (DLR), 51170 Cologne, Germany}
\author{Matthias Sperl}
\affiliation{Institut für Materialphysik im Weltraum, Deutsches Zentrum für Luft- und Raumfahrt (DLR), 51170 Cologne, Germany}
\author{Till Böhmer}
\affiliation{Institut für Materialphysik im Weltraum, Deutsches Zentrum für Luft- und Raumfahrt (DLR), 51170 Cologne, Germany}
\author{Jan Philipp Gabriel}
\affiliation{Institut für Materialphysik im Weltraum, Deutsches Zentrum für Luft- und Raumfahrt (DLR), 51170 Cologne, Germany}


\date{\today}

\begin{abstract}
Granular gases are commonly characterized through their velocity distribution, which provides access to the granular temperature. In experiments, velocity distributions are typically obtained by particle tracking, which however becomes limited at moderate and high particle densities. As a way forward, we propose a new technique for measuring particle velocities in situ by using a Fiber Bragg Grating (FBG) sensor, which remains applicable at significantly higher particle densities.The FBG sensor detects strain pulses induced by particle-fiber collisions, from which the velocity of the impacting particle can be derived. Applying this method to an ensemble of granular particles allows to extract its velocity distributions as we present for a granular system excited by a vibrational shaker. We validate the extracted velocity distribution against conventional particle-tracking measurements, confirming the reliability of the FBG-based technique. 
\end{abstract}

\maketitle


\section{\label{sec:Intro}Introduction}
Granular matter, as encountered on earth, appears both at rest and in motion - from the movement of dunes in the desert to the flow of tephra and plant seeds \cite{andreotti2013granular,battaglia2019autopsy}. By contrast, granular matter in space is rarely at rest, as seen for example in accretion disks~\cite{balbus1998instability}. This naturally raises the question: What controls the flow of granular matter? 

A central parameter is the volume fraction
\begin{equation}
    \label{eq:Volfrac}
    \phi = \frac{V_\mathrm{G}}{V_\mathrm{T}},
\end{equation}
quantifying the ratio of the volume occupied by the granular particles, $V_\mathrm{G}$ with respect to the volume they are contained in, $V_\mathrm{T}$, i.e., how densely the grains are packed~\cite{andreotti2013granular,schella2017influence,kiesgen2015vibration,schroter2017local}

However, volume fraction is not the only relevant parameter: External forces such as gravity, vibrations, shear or introducing a permeating gas or fluid flow inject energy into the granular system and thereby modify $\phi$ dynamically~\cite{andreotti2013granular,geldart1973types,duran2012sands,rosato2020segregation}. Thus, a dense granular system in rest (typically at $\phi \gtrsim\,$55\%~\cite{schroter2017local}) can be fluidized by supplying sufficient energy~\cite{rosato2020segregation}. The resulting motion depends on the excitation level: low excitations produce a rather dense fluid-like system, whereas strong excitations can create dilute gas-like systems~\cite{andreotti2013granular,duran2012sands}.

However, the transition between a granular gas and liquid remains poorly defined, showing no clear difference in governing physics and, in theory, depending only on the volume fraction~\cite{ogawa1980equations,goldhirsch2008introduction,andreotti2013granular}. Investigating this open question experimentally is complicated by the opacity of granular matter, which prevents optical techniques from following the motion of individual particles: For a dilute granular system in three-dimensional space, physical properties of individual particles, such as position and velocity, are usually obtained by using direct imaging techniques paired with particle tracking~ \cite{Aumaitre2018,Falcon1999,Hou2008,Harth2018,Steinpilz2019,Yu2019}. However, already at volume fractions as low as $\varphi=0.05$, most particles in motion are obscured by others located closed to the camera. Thus, particle tracking can not be applied straightforwardly to denser systems, where particle velocities can only be probed close to the boundary, where boundary effects are thought to dominate~\cite{YuGas,andreotti2013granular}. Approaches that can circumvent this issue, i.e. tomography techniques like CT or MRT, with a few exceptions, lack the necessary temporal resolution to access the relevant timescales of granular dynamics in such systems~\cite{lenz2000methods,stannarius2017magnetic,wang2022characterization}. Therefore trade-off between time resolution and particles perceived by a technique needs to be made as soon as the system becomes opaque.

\begin{figure}[t]
    \centering
    \includegraphics[width=0.9\linewidth]{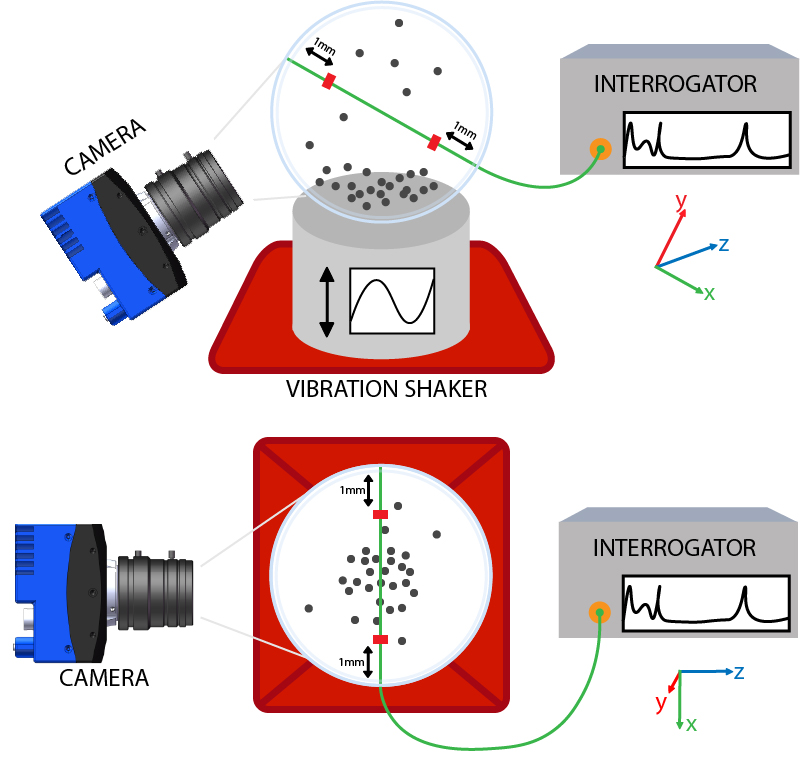}
    \caption{Schematic illustration of the employed experimental set-up, with a side-view in top and a top-view in bottom panel of the figure. The vibration shaker sinusoidally excites a spherical sample cell filled with 3\% MuMetal particles. A camera, positioned perpendicular to the fiber, records the particles in the vicinity of the fiber (green) and the FBGs (red). Collision-induced wavelengths shifts in the FBG sensors are monitored using an interrogator.}
    \label{fig:Setup}
\end{figure}

To provide a possible way forward, we introduce a method employing fiber-optical sensors based on fiber Bragg gratings (FBGs)~\cite{hill2002fiber,othonos2006fibre,jackle2019fiber}, which enables the measurement of the velocity distribution in granular systems at previously inaccessible volume fractions. The principal idea is to employ the FBG sensor to monitor the mechanical deflection of an optical fiber during collisions with granular particles. As we derive, the amplitude of a deflection can be obtained from the wavelength shift probed in the FBG sensor and provides insights into the colliding particle's initial velocity. An exemplary experimental implementation is shown schematically in Fig.~\ref{fig:Setup}, where an optical fiber containing two FBG sensors is mounted to be surrounded by granular particles excited by a vibration shaker.

A unique characteristic of granular gases is that, according to kinetic theory, their velocity distribution can be described by a Maxwell-Boltzmann distribution with overpopulated tails~\cite{YuGas,vanNoije1998,campbell1990rapid,andreotti2013granular}. Additionally, due to granular particles colliding inelastically, the restitution coefficient $\varepsilon_R$, indicating the degree of elasticity, is introduced \cite{YuGas,andreotti2013granular,haff1983grain}. Since only the exponents are of interest the distribution can be simplified to $P(v) \propto \exp(-kv^{3/2})$ for $v > \langle v\rangle/\varepsilon_R$, where $k = m/(2\,k_B\,T)$, $T$ the granular temperature, $m$ the particles' mass and $\langle v\rangle$ their average velocity. 

Below, we begin by discussing the principles of FBG sensors and the considerations required to employ them to probe a granular system's velocity distribution. Afterwards, we present the first experimental results, beginning with a careful evaluation of single-particle collisions with known initial velocities and contact angle. Finally, we employ an FBG sensor to probe the velocity distribution of a granular system at 3\% volume fraction and show that the results conform to simultaneous particle-tracking measurements and to previous experiments on granular gasses~\cite{YuGas,harth2013granular,rouyer2000velocity}.

\section{\label{sec:Methods}Experimental Principle}

\begin{figure}[t!]
    \centering
    \includegraphics[width=1\linewidth]{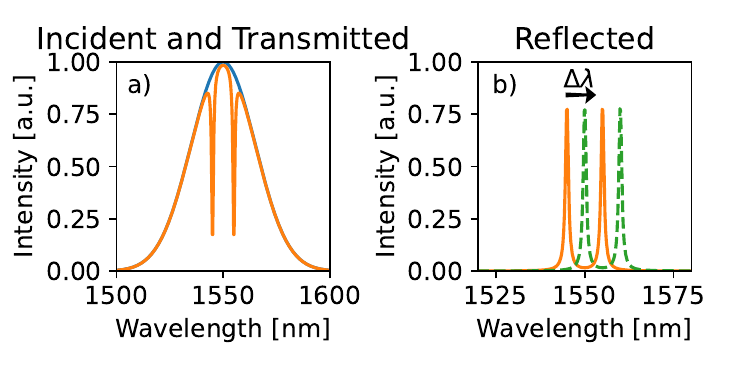}
    \caption{Schematic illustration of the incident, transmitted and reflected light's spectrum in a FBG sensor. The left panel shows the normalized incident spectrum before passing the grating (blue) and the light transmitted through the grating (orange). The reflected light, which is the one detected by the interrogator, is shown in the right panel. The reflected signal shows a wavelength shift $\Delta\lambda$ once the FBG is strained (dashed green).}
    \label{fig:OVSpectra}
\end{figure}

FBG sensors are optical sensors that use the principle of Bragg reflection to measure various physical parameters such as strain, temperature, and pressure \cite{hill2002fiber,othonos2006fibre,Bragg1913}. These sensors are produced by inscribing a periodic variation of the refractive index (e.g. by femtosecond laser technology) within the core of an optical fiber. When light is transmitted through the fiber, specific wavelengths are reflected by each segment and any changes in the reflected wavelengths can be correlated to local changes in the physical deformation of the grating. For the measurement, the fiber is connected to a fiber optical readout unit, referred to as FBG interrogator. The readout unit consists of a broadband light source, e.g., a super-luminescent diode, as well as a spectrometer or a photo-diode employed to analyze the reflected light~\cite{manavi2024deep,waltermann2014fiber}. Following this, as illustrated in Fig. \ref{fig:OVSpectra}a, the intensity spectrum of light transmitted through the FBG sensor (depicted in orange) features a valley in intensity at the Bragg wavelength compared to the incident light (blue). Thus, the reflected light contains the reflected Bragg wavelengths of the FBG sensors, as shown in Fig. \ref{fig:OVSpectra}b. When a strain is applied to the fiber, it induces a shift $\Delta\lambda$ to the reflected wavelengths (depicted in green), which is the quantity monitored as a function of time during a typical FBG measurement. As the fiber temperature in our measurements is assumed to be stationary on the time scale relevant for the experiment, changes of $\Delta\lambda$ mostly reflect changes of mechanical strain $\varepsilon$, which is the ratio of change in length of the fiber, $\Delta L$, to original length, $L$, due to a tensile force, i.e., \cite{Meyer2015}
\begin{equation}
    \label{eq:Strain}
    \varepsilon= \frac{\Delta L}{L}
\end{equation}
While neglecting temperature-induced effects, the wavelength shift, $\Delta\lambda$, caused by this strain can be calculated as:
\begin{equation}
    \label{eq:WLShift}
    \Delta\lambda = \lambda_{\mathrm{B}_0} \cdot (1-c_\mathrm{Photo}) \cdot \varepsilon,
\end{equation}
where $\lambda_{B_0}$ is the Bragg wavelength and $c_\mathrm{Photo}$ the material-dependent photo-elastic constant, which for the FBG sensor used in this study is $\sim 0.22$~\cite{Meyer2015,jackle2019fiber}. Considering the commonly available accuracy of this measurement technique of $\Delta\lambda\approx1\,$pm, strains can be detected with an accuracy of $\pm0.83\,$\textmu m/m at $\lambda_{\mathrm{B}_0} = 1555\,\mathrm{nm}$.

When the FBG sensor is implemented in a granular system, collisions of particles with the fiber lead to deflection of the latter and, thus, induce a distinct strain. In the following, we derive a relation between the velocity $v$ of the particle hitting the fiber and the observed wavelength-shift $\Delta\lambda$. To start, we consider direct collisions with vanishing contact angle and with the velocity vector perpendicular to the direction of the fiber. Deviations from these conditions are treated below.

\begin{figure}[t!]
    \centering
    \includegraphics[width=1\linewidth]{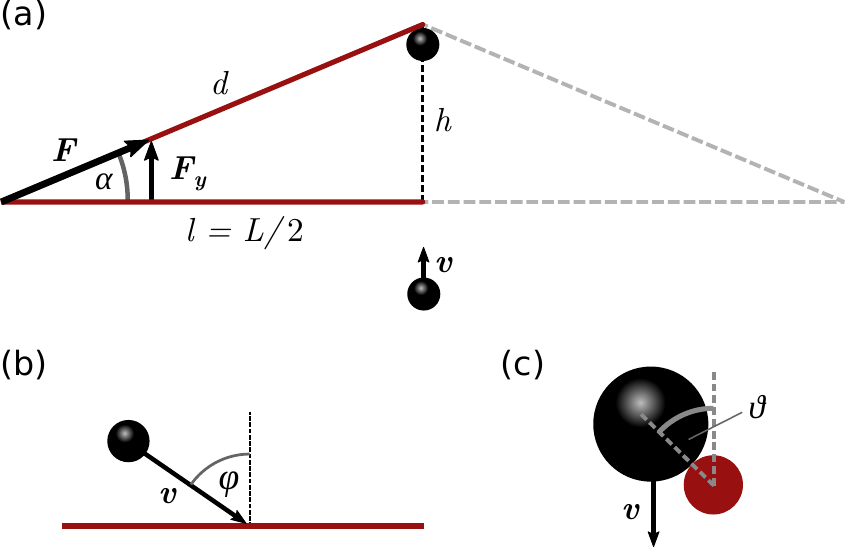}
    \caption{(a) Illustration of the fiber-deflection during particle-fiber collision, defining the parameters for the derivation of the $v(\Delta\lambda)$ relation. (b) Definition of the polar angle $\varphi$ and (c) the contact angle $\vartheta$ of particle-fiber collision.}
    \label{fig:geometry}
\end{figure}

Fig.~\ref{fig:geometry}a illustrates the deflection of the fiber after collision with a particle with velocity $v$ and mass $m$. We define the maximum deflection amplitude during the collision as $h$. Thus, neglecting dissipation, the work $W(h)$ required to deflect the fiber by $h$ equals the particle's initial kinetic energy, 
\begin{equation}
\label{eq:energy_conservation}
    W(h) = \frac{1}{2}mv^2.
\end{equation}
For simplicity, we consider a central impact, where the particle strikes the midpoint of the fiber. In this symmetric case it is sufficient to analyze one half of the fiber, i.e., one only needs to consider the left-hand triangle in Fig.~\ref{fig:geometry}a. To obtain an explicit expression for $W(h)$, we consider the absolute value of the force $F$ acting along the fiber direction, 
\begin{equation}
    F = k\,(d-l)+\frac{T}{2},
\end{equation}
where $k$ is the fiber's axial spring constant, $T$ the applied pre-tension, $l=L/2$ is half the non-deflected length and $d$ is half the deflected length.

To determine the work along the direction of $v$, we consider only the corresponding force component
\begin{equation}
\label{eq:force_res}
    F_y = F\,\sin\alpha= \frac{h}{d}\,F = kh + \frac{h}{\sqrt{h^2+l^2}}\left( \frac{T}{2}-kl\right),
\end{equation}
where $\alpha$ is the angle of deflection and the Pythagorean relation $d = \sqrt{l^2+h^2}$ has been used.

The work required to deflect the fiber by $h$ is obtained by integration, which yields
\begin{equation}
    \begin{aligned}
        \frac{W}{2} &= \int_0^h \left[ ks+\left(\frac{T}{2}-kl\right)~\frac{s}{\sqrt{s^2+l^2}}\right]~\mathrm{d}s \\
        &= \frac{k}{2}h^2 + \left(\frac{T}{2}-kl\right)\left(\sqrt{h^2+l^2}-l\right)
    \end{aligned}
\end{equation}
From Eq.~(\ref{eq:Strain}), the strain in the fiber is given by
\begin{equation}
    \varepsilon=\frac{d-l}{l}
\end{equation}
and, upon combining this relation with the energy conservation condition Eq.~(\ref{eq:energy_conservation}), we obtain
\begin{equation}
    v^2 = \frac{2l}{m}\left(kl\varepsilon^2+T\varepsilon\right)
\end{equation}
which links the particle's initial velocity to the resulting maximum strain in the fiber. Considering Eq.~(\ref{eq:WLShift}), a similar relation holds for the particle velocity and the probed wavelength shift, i.e.,
\begin{equation}
    \label{eq:vwavelength}
    v^2 = \frac{2l}{m\kappa}\left(\frac{kl}{\kappa}\Delta\lambda^2+T\Delta\lambda\right),
\end{equation}
where the constant $\kappa=\lambda_{\mathrm{B}_0}(1-c_\mathrm{Photo})$ relates strain and wavelength shift. 

For each detected collision event probed by the FBG, the derived relation allows us to determine the initial velocity of the colliding particle. Noteworthy, the relation, for now, is only valid for direct collisions with vanishing polar angle $\varphi$ and contact angle $\vartheta$ - see Fig.~\ref{fig:geometry} for an illustration of both angles. Utilizing the FBG sensor to probe the velocity distribution of a granular system requires to explicitly take into account collisions with non-vanishing $\varphi$ and $\vartheta$. Thus, in the following we treat how the distribution of particle velocities is modulated when both angles are considered.

Following arguments of classical mechanics and neglecting frictional effects, the particle's velocity component $u$ relevant for the deflection of the fiber is given by 
\begin{equation}
    u = v\,\cos(\vartheta)\cos(\varphi).
    \label{equ:reduced_velocity}
\end{equation}
When a large number of collisions from an ensemble of granular particles is probed, $v$, $\vartheta$ and $\varphi$ are random variables distributed as probability densities $P_V(v)$, $P_\Theta(\vartheta)$ and $P_\Phi(\varphi)$, respectively. $P_V$ is the velocity distribution of interest, while we determine $P_\Theta(\vartheta)$ and $P_\Phi(\varphi)$ using the following geometrical arguments:

In an isotropic granular system, no direction of particle velocity is favored, such that $\varphi$ is uniformly distributed as $P_\Phi(\varphi)=1/\pi$ for $\varphi \in [-\pi/2,\pi/2]$.

Considering translational invariance, the impact parameter $b=(R+r)\sin\vartheta$ of particle-fiber collisions is uniformly distributed for $b\in[-(R+r),R+r]$, where $R$ and $r$ are the fiber's and the particle's radius, respectively. Variable transformation yields for the contact angle distribution
\begin{equation}
    P_\Theta(\vartheta) = P_B\bigl(b(\vartheta)\bigr)\left|\frac{\mathrm{d}b}{\mathrm{d}\vartheta}\right|=\frac{1}{2}\cos(\vartheta).
\end{equation}

Combining these ideas, we obtain for the probability density $P_U(u)$ of the reduced velocity

\begin{equation}
\begin{aligned}
    P_U(u) = \frac{1}{m}
    \int\limits_{-\frac{\pi}{2}}^{\frac{\pi}{2}} \!\!
    \int\limits_{-\frac{\pi}{2}}^{\frac{\pi}{2}}
    & P_V\!\left( \frac{p}{\cos\vartheta \cos\varphi} \right)
    \frac{P_\Phi(\varphi)\,P_\Theta(\vartheta)}{\cos\vartheta \cos\varphi} \\
    & \times \mathrm{d}\varphi\,\mathrm{d}\vartheta.
\end{aligned}
\label{equ:prob_density_momentum}
\end{equation}

\begin{figure}
    \centering
    \includegraphics[width=1\linewidth]{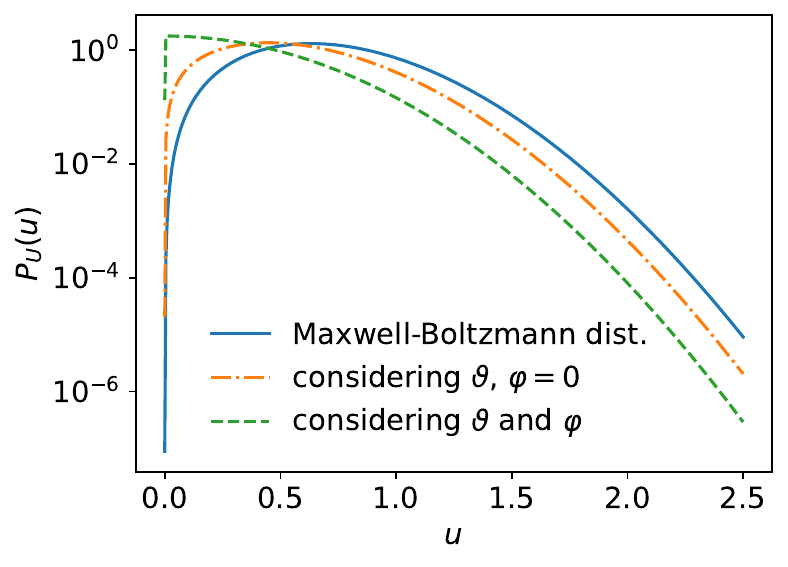}
    \caption{Velocity distribution functions for a Maxwell-Boltzmann distribution ($P_V(v_p)$) and the angle-dependent skewed versions.}
    \label{fig:prob_density}
\end{figure}

To visualize the difference between $P_U(u)$ and $P_V(v_p)$, we choose $P_V(v)$ to be Maxwell-Boltzmann distributed and calculate $P_U(u)$ numerically using Eq.~\ref{equ:prob_density_momentum}. The results are illustrated in Fig.~\ref{fig:prob_density}, where the blue and green data represent $P_V(v)$ and $P_U(u)$, respectively. In addition, we include the orange curve that assumes the polar angle to be zero, i.e., the case where particles have only velocities that are perpendicular to the fiber. Evidently, considering the distribution of contact and polar angle shifts the maximum of the probability density towards smaller velocities and distorts the shape of the velocity distribution at small velocities. However, its shape remains virtually unchanged at larger velocities, which is particularly advantageous as it enables the nature of the overpopulated high-velocity tail present in dense granular gases to be extracted.

We note that, at least in principle, $P_V(v)$ can be recovered from $P_U(u)$ by inverting Eq.~\ref{equ:prob_density_momentum}, although, in practice, such an inversion is highly sensitive to noise, and is therefore unlikely to be feasible with experimental data. However, it is possible to obtain expressions relating the mean velocities from both distributions, which reads
\begin{equation}
    \label{eq:AVGu}
    \langle u\rangle = \bigl\langle v\,\cos\vartheta\cos\varphi \bigr\rangle =  \langle v\rangle \langle\cos\vartheta\rangle\langle\cos\varphi \rangle 
\end{equation}
and allows comparison to results from particle-tracking approaches that can extract the absolute value of the velocity, without any modulation by polar and contact angle. Considering the distributions derived above, $\langle\cos\vartheta\rangle = \pi/4 = 0.785$ and $\langle\cos\varphi\rangle = 2/\pi= 0.637$.

We conclude that analyzing $P_U(u)$ instead of $P_V(v_p)$ allows to draw qualitative conclusions about the shape of the velocity distribution in granular systems, especially at large velocities, and yields a quantitative expression for the respective mean velocities.

\noindent

Finally, we briefly discuss the expected evolution of $\Delta\lambda$ as a function of time once the fiber is hit by a granular particle. The considerations for obtaining the work required to deflect the fiber can be employed to calculate the equation-of-motion for a particle-fiber collision and, thus, the temporal evolution of the wavelength shift. While the exact solution is only available numerically, for describing the probed signals we consider the analytical expression 
\begin{equation}
    \Delta\lambda(t) = \sqrt{1+A\,\sin(\omega t)^2}-1
    \label{equ:temporal}
\end{equation}
obtained by first-order Taylor expansion (see Supplementary material). Fig.~\ref{fig:ZoomedIn} shows data for the wavelength shift probed in both gratings and a fit by Eq.~(\ref{equ:temporal}). Notably, both gratings probe the same strain, thus, below we always consider the average of both gratings and analyze collision events through fits by Eq.~(\ref{equ:temporal}).


\begin{figure}[t!]
    \centering
    \includegraphics[width=1\columnwidth]{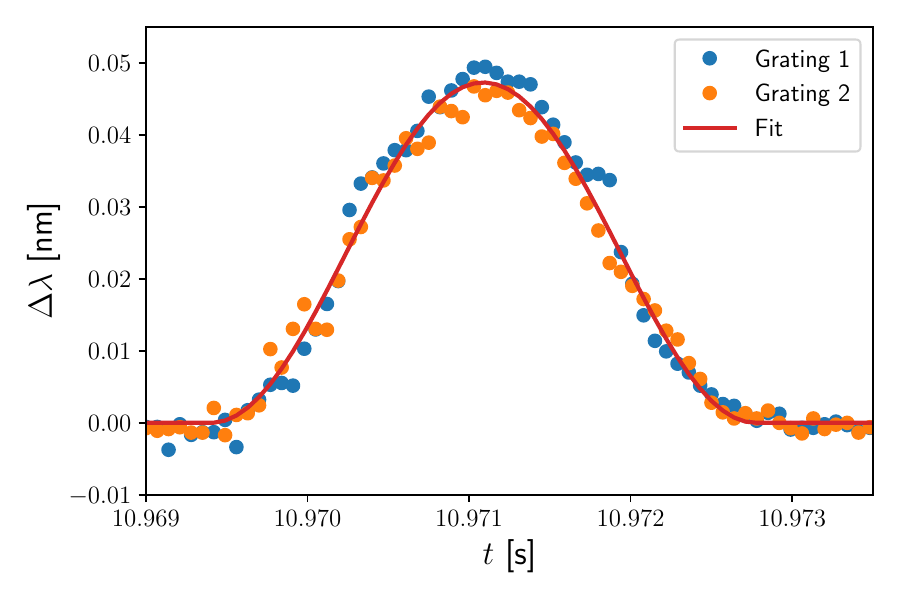}
    \caption{Wavelength shift versus time for two gratings in a 5 ms time interval. The red line is the fit by Eq.~(\ref{equ:temporal}) while the blue and orange dots represent grating 1 and 2, respectively.}
    \label{fig:ZoomedIn}
\end{figure}

\begin{figure*}[t!]
    \centering
    \includegraphics[width=\linewidth]{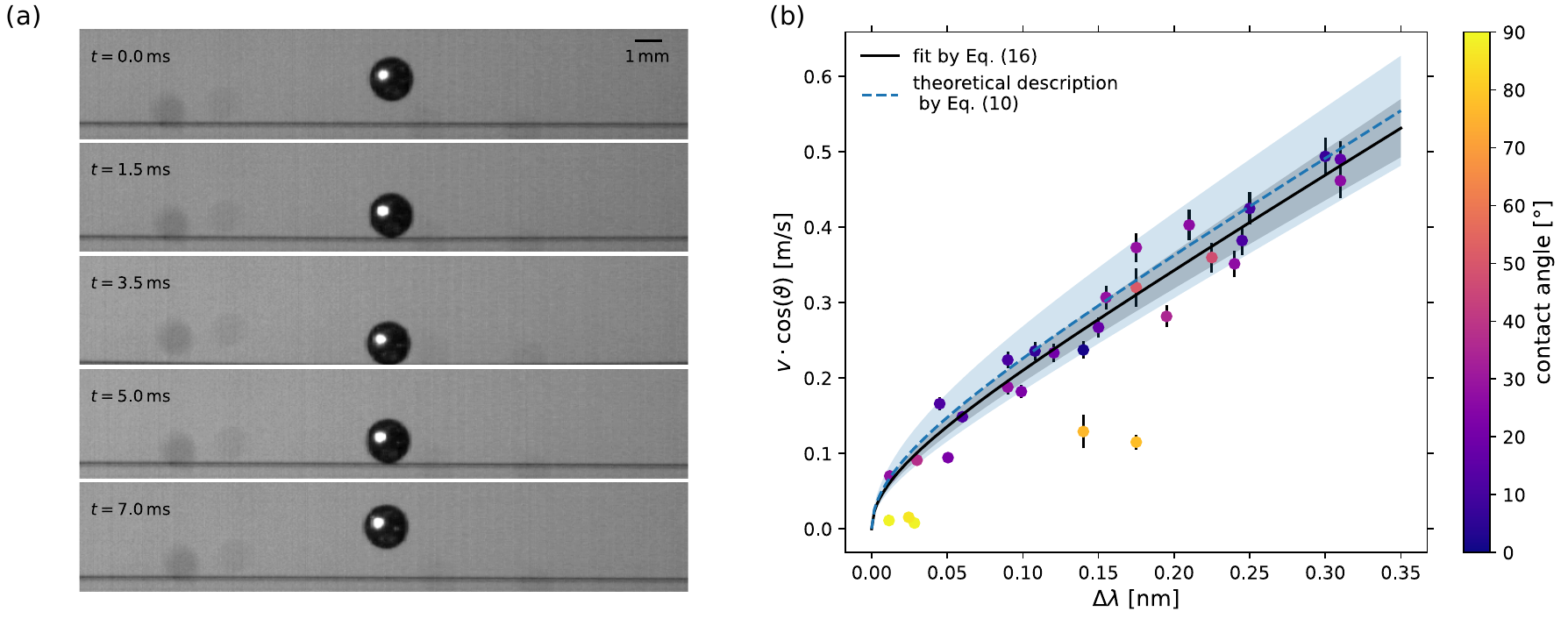}
    \caption{Interaction of a single particle with the fiber. Panel a shows an orthogonal collision event of the particle and the fiber. The collision takes roughly 3.5~ms from frame two to four. Panel b shows the reduced velocity $u=v\cos\vartheta$ of the particles plotted against wavelength shift $\Delta\lambda$ observed with the FBG. The contact angle of each data point is indicated by its color. The solid black line indicates a fit based on Eq.~(\ref{eq:fit}), while the dashed blue line is Eq.~(\ref{eq:fit}) using the experimental parameters (see text).}
    \label{fig:Lincorr}
\end{figure*}

\section{\label{sec:ExpPro}Experimental results}

To confirm the applicability of FBG sensors for measuring granular velocity distributions, we conduct two type of benchmark experiments: First, we verify the theoretically derived relationship between a particle's initial velocity and the corresponding FBG wavelength shift $\Delta\lambda$ derived in Eq.~(\ref{eq:vwavelength}). This is achieved by monitoring individual particle-fiber collisions using a high-speed camera to extract initial velocity and contact angle, and correlating the result to the amplitude of the FBG signal. The results allows us to calibrate the FBG sensor for determining absolute particle velocities. Second, we measure the velocity distribution of an ensemble of granular particles excited by a vibration shaker using the FBG signal, and confirm that it agrees with the distribution obtained via particle tracking.

Both experiments were conducted inside a sample cell consisting of a glass sphere with a radius of $r=2\,$cm. The glass fiber (SMF-28 Ultra 200 Optical Fiber from Corning, with a Young's modulus of 8~GPa, a cladding diameter of $d_\mathrm{clad} = 125\,$µm and a coating diameter $d_\mathrm{coat} = 250\,$µm~\cite{morgese2023stress,jackle2019fiber,cherkasova2025measurement}), equipped with two FBG sensors, is inserted into the cell through two small openings. The fiber was positioned such that the distance between each FBG and the corresponding cell wall were approximately 1\,mm (marked red in Fig.~\ref{fig:Setup}), ensuring that collisions between fiber and particles occurred only between the two gratings. Next, the fiber was glued to one side of the cell, after which a weight of $ 50\cdot10^{-3}\,\rm{kg}$ was attached to the fiber to apply a fixed tension force $T$. While the weight remained attached, the fiber was glued to the opposite opening. After curing the glue and removing the weight, the relaxation of the glue most likely led to a reduction of the tension to about half its initial value, as will be discussed below. The fiber was connected to the Interrogator (model SPK-155 from Miopas) operated at $1/7\cdot 10^5$ Hz and set to monitor both gratings simultaneously. We note that using a fiber with polyimide coating is necessary for these types of measurements as acrylate coating created artifacts in the signal. These were most likely caused by the core moving against the coating, as acrylate coating is less firmly attached to the core than polyimide.

In both experiments, Mu-metal particles with a diameter of $1.6\cdot10^{-3}\,\rm{m}$, a density of $8.7\cdot10^{3} \,\rm{kg/m^3}$, and a mass per particle of $1.87\cdot10^{-5}\,\rm{kg}$ were used . These were chosen to match previous experiments on magnetically agitated granular gasses in weightlessness~\cite{Yu2019}.

\subsection{\label{subsec:LinRes} Single Particle Drops}

To verify the relationship between $\Delta\lambda$ and $u$ [see Eq.~(\ref{eq:vwavelength})], single particles were dropped onto the fiber from various different heights, while their velocity $v$ and contact angle $\vartheta$ were monitored using cameras. As dropping the particles from above ensures their velocity to be perpendicular to the orientation of the fiber, the polar angle $\varphi$ is zero and was not monitored during the experiment. To obtain accurate velocity data, a MotionBLITZ Cube 4 high-speed camera, operated at 10,000 frames per second, was used to track the particles' position relative to the fiber. The contact angle was obtained as half the angle of deflection of the particles after collision with the fiber, which was monitored using a Basler acA800-510um camera operated at 510 frames per second and positioned to monitor the particles' position within the scattering plane. Exemplary images showing a particle during collision with the fiber are shown in Fig.~\ref{fig:Lincorr}a.

The particle velocities were determined using particle tracking and ranged from 0.07 to 0.54\,m$/$s with contact angles between 2 and 88\,$^\circ$. Fig.~\ref{fig:Lincorr}b summarizes the results from the single-particle drop experiments by plotting the particles' reduced initial velocity $u = v\cos(\vartheta)$ against $\Delta\lambda$. The color of each symbol indicates the contact angle according to the color bar, while the error bars represent the uncertainty of the measured velocity due to the camera resolution.

Inspired by Eq.~(\ref{eq:vwavelength}), the solid black line shows a fit using the expression
\begin{equation}
\label{eq:fit}
    u(\Delta\lambda) = \sqrt{a\, \Delta\lambda^2 + b\,\Delta\lambda},
\end{equation}
which yields $a=(1.461\pm0.295)\,\mathrm{s}^{-2}$ and $b=(0.295\pm0.060)\,\mathrm{m}\cdot\mathrm{s}^{-2}$. The fit describes the data well, except for collisions at the largest contact angles, where frictional effects rather than momentum transfer likely dominate the particle-fiber interaction. In the following, Eq.~(\ref{eq:fit}) with the parameters determined by the fit is used to extract the absolute particle velocity from $\Delta\lambda$.

For comparison, the dashed blue line represents Eq.~(\ref{eq:fit}), but with parameters $a$ and $b$ calculated from the experimental parameters according to Eq.~(\ref{eq:vwavelength}). Here, a value of $k=(45.9\pm4.6)$\,kN/m is used for the fiber's effective spring constant as a composite material with Young's moduli of 68.56~GPa and 2.5~GPa for silica and polyimide, respectively \cite{cherkasova2025measurement,Matmake_Polyimide_Properties_2025}. We calculate the spring constant instead of determining it experimentally to avoid possible complications when measuring a composite material's Young's modulus by tensile tests, as it was done in literature for fiber materials~\cite{cherkasova2025measurement,morgese2023stress,etde_21001297,hwang2019effects}. As stated above, the resulting tension of the fiber after removing the weight and relaxation of the glue is unknown. The model corresponds to the data when we assume a tension of $T=(200\pm100)\,$N, which corresponds to approximately half the tension introduced initially. Due to the uncertainty of the tension, we recommend performing an empiric calibration of the velocity-wavelength shift relationship as shown in Fig.~\ref{fig:Lincorr}. In the future, a more controlled procedure for implementing tension to the fiber could eliminate the need for such a calibration.

\subsection{\label{subsec:ShakerExp}Shaker Experiment}

\begin{figure}
    \centering
    \includegraphics[width=1\linewidth]{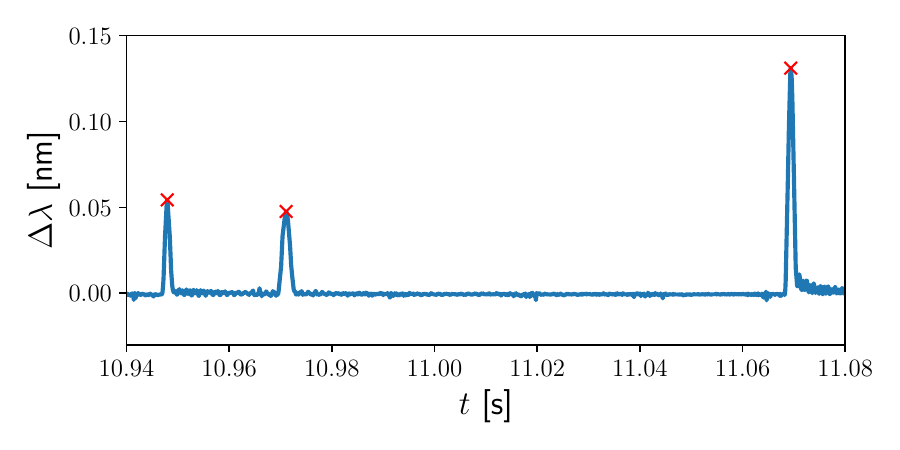}
    \caption{Wavelength shift versus time for the two averaged gratings and three collision events in a 140 ms time interval shown in blue. The red symbols indicate a detected collision event. The subtle background noise is caused by the 40\,Hz excitation the shaker operates at.}
    \label{fig:ZoomedOut}
\end{figure}

The experimental setup employed for the shaker experiment is depicted schematically in Fig. \ref{fig:Setup}. The particles were filled into into the sample cell until a volume fraction of 3\% was reached. The shaker, model LSD V721 by Brüel\&Krear, is operated at a sinusoidal frequency of 40\, Hz and an amplitude of 1\,mm is used. Data were collected for a total of 30\,min to reach sufficient statistics. 

An exemplary time series of the FBG signal is shown in Fig.~\ref{fig:ZoomedOut}, displaying three distinct collision events marked by the red symbols. For further analysis the raw signal is first filtered by applying a Gaussian filter to remove high-frequency noise. Subsequently, peak detection was performed using the \texttt{signal.find\_peaks} function from the SciPy library~\cite{virtanen2020scipy}. We only include peaks which appear in both gratings, have amplitudes larger than 0.01\,nm and a distance of at least 0.003\,s from another collision event. Subsequently, the peaks are fitted using Eq.~(\ref{equ:temporal}) to extract their amplitude $\Delta\lambda$ (see Fig.~\ref{fig:ZoomedIn}). Using Eq.~(\ref{eq:fit}), the particle's reduced velocity is calculated for each impact, which finally yields the velocity distribution of the granular system.

\begin{figure*}[t!]
    \includegraphics[width=1\linewidth]{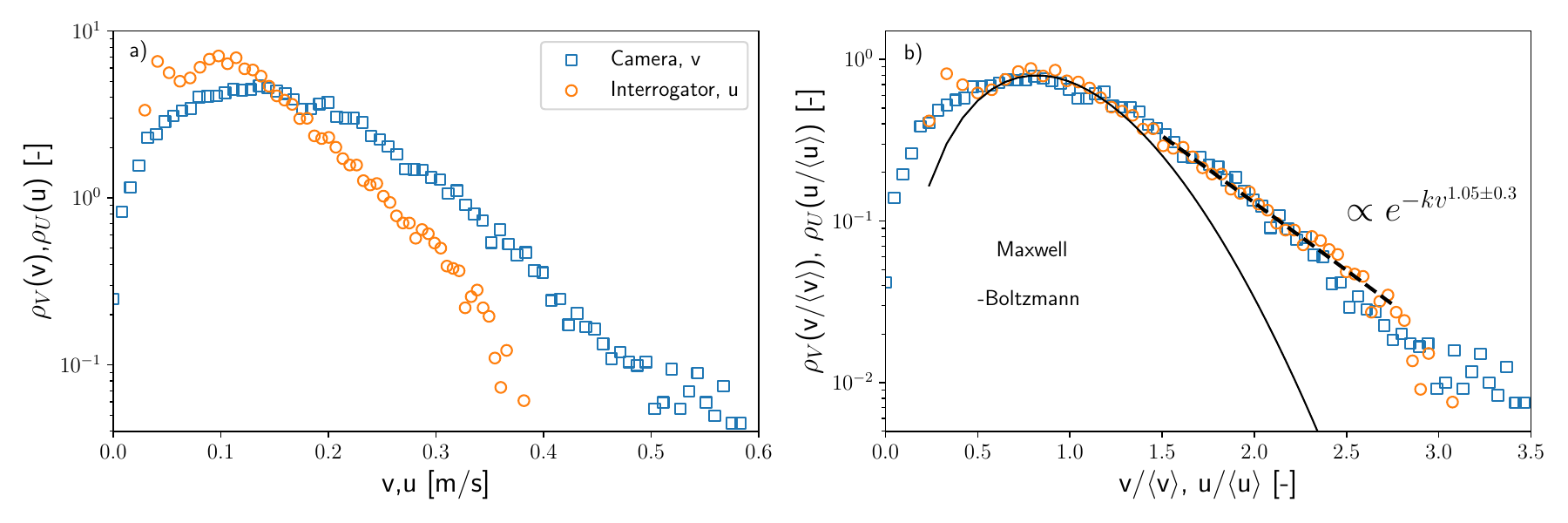}
    \caption{Velocity distribution of the recorded impacts via FBGs and particle tracking in absolute, panel (a), and normalized, panel (b), velocity. The orange circles show FBG data and blue squares show particle tracking data. The solid black line is a Maxwell-Boltzmann fit for low velocities and the dashed black line an exponential fit with an exponent of $1.05\pm0.3$, indicative of the overpopulation at high velocities.}
    \label{fig:Vels}
\end{figure*}

To compare the obtained velocity distribution to results obtained from an established technique, we also determine the velocity distribution from the video recordings of the shaker experiment (with 5000 frames per second) using particle tracking. We note that the particle tracking data correspond to an experiment shaken at 50\,Hz.

We compare the velocity distributions obtained using the FBG sensor (orange symbols) and particle tracking (blue symbols) in Fig~\ref{fig:Vels}. Panel (a) reports the distributions as functions of the absolute (reduced) velocity, while panel (b) shows them normalized with regard to their respective average velocities, which correspond to $\langle u_\mathrm{int}\rangle = 0.1244\,$m/s and $\langle v_\mathrm{PT}\rangle = 0.2143\,$m/s, respectively. We note that $\langle v_\mathrm{PT}\rangle$ includes a factor of $4/\pi$ that compensates for the fact that the technique probes a 2D projection of the 3D particle dynamics.

The shapes of both velocity distributions are very similar, which is emphasized when comparing both distributions as a function of the normalized velocities, showing almost perfect collapse. As functions of the absolute velocities, the velocity distribution obtained from the FBG sensor is shifted to lower velocities, as it is expected considering that the technique probes $v\cdot\cos\vartheta\cos\varphi$ instead of $v$. Comparing the quotient of the respective average velocities yields
\begin{equation}
    \frac{\langle u_\mathrm{int}\rangle }{\langle v_\mathrm{PT}\rangle} = 0.581,
\end{equation}
which is slightly larger than the factor of $\langle\cos\vartheta\rangle\langle\cos\varphi\rangle=0.5$ derived in Eq.~(\ref{eq:AVGu}). This is expected though, considering that collisions at large polar and contact angles close to $\pi/2$ induce very small $\Delta\lambda$ and therefore tend to fall below the detection threshold of the technique. As a result, $\langle\cos\varphi\rangle\langle\cos\vartheta\rangle$ is expected to be slightly larger than what would be expected for an isotropic velocity distribution from theory.

\section{\label{sec:Disc}Discussion}

We begin by discussing the conclusions that can be drawn from the shape of the velocity distributions presented in Fig.~\ref{fig:Vels}b. Both employed experimental techniques reveal an overpopulation at large velocities, which becomes evident when comparing the experimental results to the solid black curve representing a Maxwell-Boltzmann distribution that is unable to describe the data at larger-than-average velocities. Instead, the data follow the dashed black line, which is a fit to the slope of the overpopulated high-velocity tail by $\propto\exp(B\cdot v^\alpha)$, where the exponent is $\alpha=1.05\pm0.3$. This exponent suggests the studied granular system to behave like a freely cooling granular gas ($\alpha = 1$). However, due to the relatively large uncertainty of $\alpha$, we cannot fully rule out heated granular gas or even granular fluid-like behavior, which would correspond to values of $\alpha$ between 1.5 and 2  \cite{rouyer2000velocity,huan2004nmr, brey1999high}.

A precise characterization of the studied granular system's nature is further complicated considering that excitation with a shaker under gravity conditions does not create a true granular gas, due to introducing anisotropic particle velocities~\cite{brilliantov2010kinetic,falcon2013equation,rouyer2000velocity,andreotti2013granular}. 
Additionally, using a shaker for exciting the entire sample cell introduces motion of the fiber, such that the fiber probes particle velocities in the moving reference frame of the sample cell, instead of a fixed reference frame. 
Both shaker-related issues could be resolved by using another mechanism of excitation in a microgravity environment as done, e.g., by Yu et. al.~\cite{YuGas} and Sata et. al.~\cite{sata2025criteria} during parabolic flights. Implementing the FBG technique into similar experimental scenarios in the future is expected to unfold its full potential. 


Finally, we discuss the applicability of the FBG sensor for probing velocity distributions in a more dense granular system. At the studied volume fraction of 3\%, 10214 collision events were probed by the FBG sensor during the 30\,min of experiment, which corresponds to 5.6 collisions per second. Considering that the duration of a single collision is approx. 3-4\,ms (see Fig. \ref{fig:TimeDistri}), it could be possible to resolve up to 250 collision events per second, which implies that the FBG sensor remains applicable for investigating granular systems at significantly larger volume fractions. By contrast, particle-tracking techniques have been shown to struggle with the overlapping of particles in the camera recording at packing fractions beyond 5\%~\cite{YuGas}. While camera particle tracking is thus limited to mostly boundary effects for more than $\varphi = 5\%$, FBG sensors can resolve bulk behavior, since signals from all along the fiber are recorded. Experiments performed with particle tracking in conjunction with FBGs should be able to record boundary and bulk behavior at desired time scales, thereby supplementing each other well. This gives FBGs a significant advantage over tomography techniques, which operate at low frequencies and are usually spatially demanding, requiring large experimental set-ups. Furthermore their tracking data, thousands of frames, difficult to save and store would be reduced to a velocity distribution~ \cite{lenz2000methods,stannarius2017magnetic,wang2022characterization}. The required duration of a measurement to extract sufficiently statistically backed up velocity distributions with FBGs at below and slightly above  $\varphi = 3\%$ is longer than for particle tracking. However, since the measurement time scales with the collision frequency, the measurement duration should decrease with increasing density. Strengthening the application for high volume fractions.

\section{\label{sec:Conclusion}Conclusion}

We introduced an experimental technique for determining velocity distributions of granular systems based on monitoring the deflection of an optical fiber equipped with FBG sensors upon collisions with granular particles. Using classical mechanics, we derived an expression for the strain-induced wavelength shift $\Delta\lambda$ induced in the FBG when the fiber collides with a granular particle with velocity $v$. With this, evaluating the distribution of $\Delta\lambda$ obtained by analyzing numerous collision events allows to determine the velocity distribution of the granular system - which on top requires to consider how it is modulated by the polar- and contact-angle distribution of the collisions.

By performing controlled particle-fiber collisions monitored by a high-speed camera, we experimentally confirmed the derived relation between $\Delta\lambda$ and $v$ and obtain a quantitative calibration to extract absolute particle velocities from the FBG signal. 

Building on this, we applied the technique to a granular ensemble excited by a vibrational shaker. Parallel particle-tracking experiments allowed us to confirm that both techniques effectively probe the same velocity distribution. The extracted velocity distribution displays characteristic features of a freely cooling granular gas, namely an exponential deviation from Maxwell-Boltzmann statistic at high velocities with an exponential exponent of $1.05\pm0.3$, thus resembling previous results~\cite{YuGas,huan2004nmr,falcon2013equation}. Finally, we inferred that the technique is applicable to granular systems at higher volume fractions where particle-tracking techniques start to fail due to particles overlapping in the recorded videos.

\begin{acknowledgments}
We wish to acknowledge the support of  Friedrich Bürger, Tobias Schossig, Karsten Tell and Felicitats Bujnoch. Furthermore, we thank Jens Kosmann and the Department of Composite Design from DLR Braunschweig for their aid in our Interrogator problem. This work was supported by the DLR Space Administration with funds provided by the Federal Ministry for Economic Affairs and Climate Action (BMWK) based on a decision of the German Federal Parliament under Grant No. 50WM1945 (SoMaDy2).
\end{acknowledgments}

\section*{DATA AVAILABILITY}
\noindent
The data that support the findings of this article are not
publicly available. The data are available from the authors
upon reasonable request.
\bibliography{MEGraMa}

\clearpage


\twocolumngrid

\renewcommand\thefigure{A\arabic{figure}} 
\renewcommand\thetable{A\arabic{table}} 

\renewcommand{\theequation}{A\arabic{equation}}
\setcounter{equation}{0}
\setcounter{figure}{0}

\appendix
\section{Equation of Motion}
\label{App:EqMotion}
\noindent
To characterize the time evolution of the fiber, we study the differential equation given by the forces acting on the particle (Eq. \ref{eq:force_res}). To derive an analytic solution, we approximate the restitution force by retaining only the linear term in $h$. This is justified because $h<<l$.

\begin{equation}
    \begin{aligned}
    m\ddot{h} = F &= 2 \left(kh + \left( \frac{T}{2}-kl\right)\frac{h}{\sqrt{h^2+l^2}}\right) \\
    &=2kh + \left( T-2kl\right)\left( \frac{h}{l} - \frac{h^3}{2l^3} + \dots\right) \\
    &= \frac{T}{l}h +\mathcal{O}(h^3)
    \end{aligned}
\end{equation}

The fiber impact may therefore be approximated with the harmonic oscillation $h(t)= \frac{v_0}{\sqrt{T/lm}}\sin\left(\sqrt{\frac{T}{lm}}t\right)$

\noindent
Using equations \eqref{eq:Strain},\eqref{eq:WLShift} for the strain and wavelength shift the measured signal should have the form
\begin{equation}
    \Delta\lambda(t) = \kappa  \left(\sqrt{1+\frac{A^2}{l^2}\sin^2(\omega t)}-1\right)
\end{equation}
where $A=\frac{v_0}{\sqrt{T/lm}}$ and $\omega=\sqrt{\frac{T}{lm}}$. Following this the linear approximation shows no velocity dependence of the contact times, and enables us to study the tension of the fiber. Higher order terms show a strong velocity dependence of the contact time

\nocite{*}

\section{A. Contact Time distribution}
\begin{figure}[th!]
    \centering
    \includegraphics[width=1\linewidth]{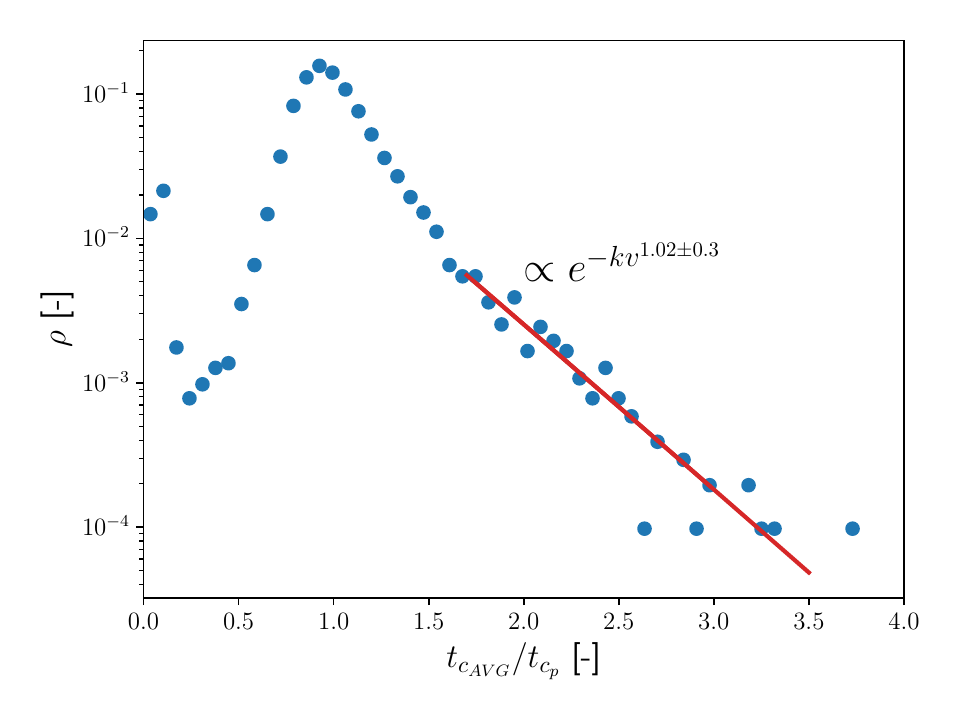}
    \caption{Contact time distribution of the recorded impacts via FBGs. The solid red line is an exponential fit with an exponent of $1.02\pm0.3$, at low impact velocities.}
    \label{fig:TimeDistri}
\end{figure}

The distribution of contact times is normalized and shown in Fig.~\ref{fig:TimeDistri}. We can not apply the same statistics to the time we used for the velocities. Furthermore low values in normalized time correspond to high velocities.
\section{Wavelength shift vs. Contact Time}
\begin{figure}[th!]
    \centering
    \includegraphics[width=1\linewidth]{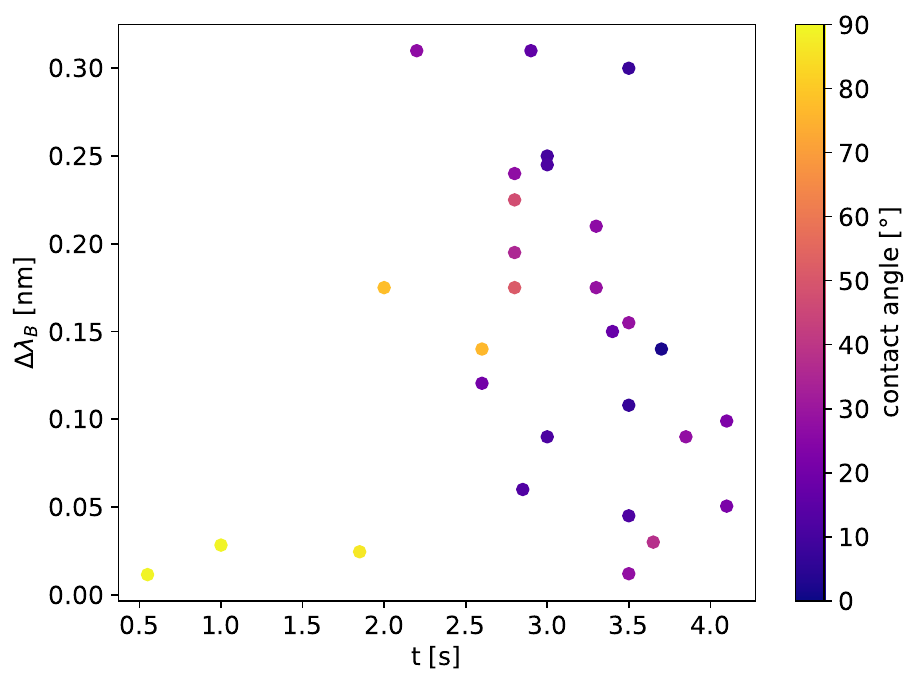}
    \caption{The contact time of the particles plotted against wavelength shift $\Delta\lambda$ measured by the interrogator. The angle dependence is indicated by the color of each symbol.}
    \label{fig:AmpTimeCorr}
\end{figure}

The distribution of contact times for the single particle drop experiment is shown in Fig.~\ref{fig:AmpTimeCorr}.

\clearpage

\end{document}